\documentclass[conference,a4paper]{APSIPA2025}
\usepackage{amsmath}
\usepackage{graphicx}
\usepackage{multirow}
\usepackage{threeparttable}
\usepackage[backend=biber,style=ieee,]{biblatex}

\usepackage{amssymb}
\usepackage{amsfonts}
\usepackage{algorithmic}
\usepackage{array}
\usepackage{tablefootnote}
\usepackage{textcomp}
\usepackage{balance}
\usepackage{bbm}
\usepackage{xcolor}

\usepackage{geometry}
\usepackage{fancyhdr}

\fancypagestyle{firststyle}{
  \fancyhf{}
  \fancyhead[C]{2025 Asia Pacific Signal and Information Processing Association Annual Summit and Conference (APSIPA ASC)}
}

\begin{document}

\title{Voice Conversion Augmentation for \\ Speaker Recognition on Defective Datasets}

\author{
\authorblockN{
Ruijie Tao\authorrefmark{1}, 
Zhan Shi\authorrefmark{3}\authorrefmark{2}, 
Yidi Jiang\authorrefmark{1}, 
Tianchi Liu\authorrefmark{1}, 
and Haizhou Li\authorrefmark{1}\authorrefmark{3}\authorrefmark{4}
}

\authorblockA{
\authorrefmark{1}
National University of Singapore, Electrical and Computer Engineering, Singapore
}

\authorblockA{
\authorrefmark{2}
University of Houston, C.T. Bauer College of Business, USA
}

\authorblockA{
\authorrefmark{3}
The Chinese University of Hong Kong, School of Artificial Intelligence, Shenzhen, China
}

\authorblockA{
\authorrefmark{4}
Shenzhen Research Institute of Big Data, Shenzhen, China
}
}

\maketitle
\thispagestyle{firststyle}
\pagestyle{fancy}

\begin{abstract}
Modern speaker recognition system relies on abundant and balanced datasets for classification training. However, diverse defective datasets, such as partially-labelled, small-scale, and imbalanced datasets, are common in real-world applications. Previous works usually studied specific solutions for each scenario from the algorithm perspective. However, the root cause of these problems lies in dataset imperfections. To address these challenges with a unified solution, we propose the Voice Conversion Augmentation (VCA) strategy to obtain pseudo speech from the training set. Furthermore, to guarantee generation quality, we designed the VCA-NN~(nearest neighbours) strategy to select source speech from utterances that are close to the target speech in the representation space. Our experimental results on three created datasets demonstrated that VCA-NN effectively mitigates these dataset problems, which provides a new direction for handling the speaker recognition problems from the data aspect.
\end{abstract}

\section{Introduction}
Speaker recognition (SR) aims to verify the identity of a speaker based on the voice characteristic~\cite{Tomi, Reynolds2000, Two_decades}. In recent years, deep learning-based SR systems have achieved significant improvement~\cite{desplanques2020ecapa, wang2023cam++, xvectors, liu2022neural}. These modern solutions typically employ a classification learning pipeline, relying on datasets with abundant labelled utterances and a balanced distribution~\cite{Voxceleb2, Voxceleb, SITW_mitch}. However, in real-world applications, dataset collection and speech labelling can be challenging due to diverse issues, such as minor languages, limited labor cost and data source~\cite{liu2024disentangling, tao2021self, han2023self}. That leads to the defective SR datasets.

We take three typical SR dataset problems for consideration. Firstly, in the partially labelled datasets, semi-supervised SR has the problem of leveraging unlabelled utterances effectively~\cite{inoue2020semi}. Secondly, for the small-scale dataset, it is challenging to learn a generalized model from limited data~\cite{li2020automatic}. Lastly, for the dataset with an imbalanced speaker distribution, the learnt model tends to prioritize speakers who contribute the majority of utterances, while ignoring those with fewer utterances~\cite{karthikeyan2021strong}.

Previous studies usually tackled each issue with a specific method. For instance, contrastive learning~\cite{inoue2020semi} and curriculum learning~\cite{zheng2019autoencoder} are studied to address semi-supervised SR. \cite{li2020automatic} proposed an adversarial few-shot learning method for the SR dataset with limited data. However, these solutions primarily focus on algorithmic aspects rather than directly addressing dataset deficiencies. Recent work \cite{wubet2022voice} has demonstrated that data augmentation with generation pipelines is also a feasible approach for speech keyword detection. \cite{Stargan2023} and \cite{huang2021synth2aug} proved that text-to-speech (TTS) generation technology~\cite{popov2021grad, huang2021unit} can be used to assist cross-domain speaker verification. Nonetheless, it remains unexplored to ensure the quality of generated utterance and provide a unified solution for all the aforementioned problems.

For these defective datasets, every utterance in the training set shares the different quality and importance~\cite{song2022learning, natarajan2013learning}. In other words, augmenting the dataset by making these imperfect utterances learn from the valuable utterances is an efficient direction. Motivated by that, voice conversion (VC) technique is a suitable choice to supplement the dataset. Here, VC aims to change the voice characteristic of one source speech based on the target speech while preserving linguistic information~\cite{sisman2020overview, mohammadi2017overview, zhou2019cross}. \cite{qin2022vc} proposed VC for speaker verification augmentation for the text-dependent condition but ignored the importance of the quality of the generated data.

In this paper, we propose the Voice Conversion Augmentation (VCA) strategy to address the aforementioned SR problems from the dataset perspective. It intelligently selects source and target speech from the training set, and generates the pseudo speech to involve diverse voiceprint features for augmentation. Meanwhile, to enhance the quality of the generated utterances, we propose a VCA-NN (nearest neighbours) strategy to select appropriate source speech that has the similar representation to the target speech during the VCA process. 

The contribution of this paper can be summarized as follows: 1) We propose VCA, a novel and unified augmentation strategy based on voice conversion to handle diverse dataset problems in speaker recognition. 2) We design the VCA-NN strategy to search for the optimal target and source speech to generate high-quality pseudo speech. 3) We create three datasets to represent scenarios of semi-supervised, small-scale, and imbalanced learning to validate the effectiveness of VCA.

\begin{figure*}[!t]
    \centering
    \includegraphics[width=.9\linewidth]{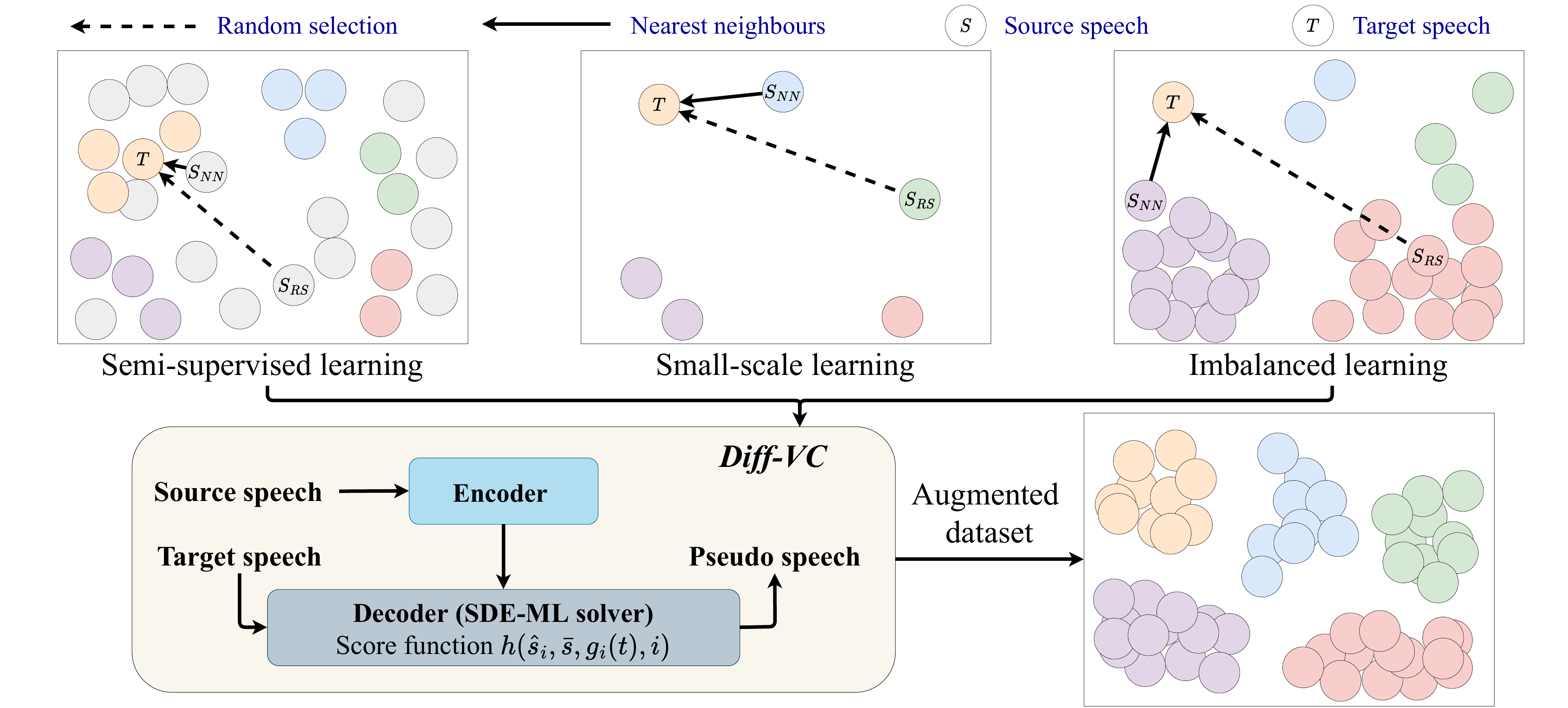}
    \caption{Data distribution illustration for three kinds of typical defective speaker recognition datasets and the augmented dataset with the Voice Conversion Augmentation (VCA) strategy. For augmentation, the Diff-VC module transfers the source speech into the pseudo speech with the target speech's voice characteristic. In data distribution illustration, circles with the same colour represent the utterances from the same speaker, and the grey circles denote the unlabelled utterances. For each target speech, VCA-RS selects random utterance as the source speech, while VCA-NN searches nearest neighbours to boost the quality of the pseudo speech.}
    \label{fig:VC}
\end{figure*}
\vspace{4mm}
\section{Background: Diffusion-based voice conversion}
Voice conversion (VC) transforms the source speech $s$ into the new speech, i.e., `pseudo speech', preserving the linguistic content of $s$ and the speaker characteristic of the target speech $t$. Based on that, as an augmentation proposal, VC can involve multiple speech representations for the same person with different prosody, content, and emotions. Our proposed VCA employs the diffusion-based voice conversion (Diff-VC) method~\cite{popov2021diffusion} due to its stable performance.

Diff-VC follows the idea of the auto-encoder. The encoder is designed to capture speaker-independent speech characteristics. It transforms the mel-spectrogram of $s$ into a speaker-independent representation, denoted as $\bar{s}$, which is further fed into the decoder. The decoder is tasked with the reverse diffusion process~\cite{ho2020denoising}, which aims to convert the speech representation into a new mel-spectrogram. This output reflects the voiceprint of the target speaker instead of the source speaker. It is formalized through a reverse stochastic differential equation (SDE)~\cite{kloeden1992stochastic}, given by:
\begin{equation}
d\hat{s}_i = \left( \frac{1}{2}(\bar{s} - \hat{s}_i) - h(\hat{s}_i, \bar{s}, i) \right)\beta_{i}di + \sqrt{\beta_{i}}d\overleftarrow{W}_i,
\end{equation}
where $\hat{s}_i$ represents the evolving speech during reverse diffusion at iteration time $i$, with $\bar{s}$ as the encoder-derived representation. The score function \(h(\hat{s}_i, \bar{s}, i)\) is trained to incorporate the target speaker's characteristic $g_i(t)$. In addition, $\beta_i$ modulates the noise level, and $d\overleftarrow{W}_i$ adds randomness through the backward Wiener process in SDEs. The resultant mel-spectrogram can closely mirror the target speaker's voice, and it can be recovered into the 1D speech waveform as the pseudo speech.

\section{Voice conversion augmentation (VCA)}

As shown in Figure~\ref{fig:VC}, VCA employs the voice conversion technique to synthesize augmented data and mitigate three typical dataset problems. Both the source speech set $\mathcal{S}$ and target speech set $\mathcal{T}$ come from the training set $\mathcal{R}$. 

\subsection{Semi-supervised learning}
In the semi-supervised learning scenario, $\mathcal{R}$ consists of $N$ labelled utterances to form the target speech set in VCA, denoted as $\mathcal{T} = \{(x_i,y_i)\}_{i=1}^{N}$, while unlabelled utterances form the source speech set in VCA, denoted as $\mathcal{S} = \{(x_j)\}_{j={N+1}}^{|\mathcal{R}|}$ ($y_j$ is unknown). The training challenge arises since unlabelled utterances cannot directly contribute to the classification learning. To leverage the unlabelled utterances effectively, we implement voice conversion for each utterance in the labelled target set $\mathcal{T}$, selecting source speech from the unlabelled set $\mathcal{S}$. After conversion, the synthetic utterance exhibits similar voice characteristic to the target speaker, while the speech content and prosody differ. This process is repeated $K$ times for each target speech by selecting the different source speech. $K$ is named as `generation coefficient', which means the number of augmented samples for each speech utterance. The augmented dataset provides a more comprehensive speech representation for each speaker, which can benefit the SR training.

\subsection{Small-scale learning}
In the small-scale learning scenario, $\mathcal{R}=\{(x_i,y_i)\}_{i=1}^{{|\mathcal{R}|}}$, where $|\mathcal{R}|$ is small. The limited dataset size can adversely affect the generalizability of the learned speaker model. In VCA, $\mathcal{R}$ plays the role of the source speech set $\mathcal{S}$ and the target speech set $\mathcal{T}$ at the same time. For each sample in $\mathcal{T}$, we treat it as the target utterance and perform voice conversion with the sample from the source set $\mathcal{S}$. This process is repeated $K$ times to achieve various speaker representations. The target and source speech should come from different speakers to guarantee the diversity of the generated utterances.

\subsection{Imbalanced learning}
The imbalanced dataset contains the minority part and the majority part. The minority part has $M$ utterances and each speaker in it has a small number of utterances, this part be viewed as the target speech set $\mathcal{T} = \{(x_i,y_i)\}_{i=1}^{M}$ in VCA. On the other hand, each speaker from the majority part has a large number of utterances, used as the source speech set $\mathcal{S} = \{(x_j,y_j)\}_{j={M+1}}^{|\mathcal{R}|}$. This imbalanced distribution makes the trained SR model more biased towards the speakers in the majority part, resulting in poor representation learning. For each target utterance of minority part, VCA is applied $K$ times to generate several pseudo utterances for supplementary. This process can achieve a more balanced speaker distribution to enhance the SR system.

\section{Nearest neighbours selection}
\label{sec}
To study the strategies of selecting the optimal utterance as the source speech, we take semi-supervised learning as an example, and other problems can apply analogous methods.

For each target speech, a straightforward strategy is \textbf{VCA-RS (random selection)}, which randomly chooses one sample from the entire source speech set $\mathcal{S}$ as the source speech (represented as $S_{RS}$ in Figure~\ref{fig:VC}). However, VCA-RS lacks control over the quality of the generated utterance. For instance, transferring an utterance from a female's voice to a male's voice is challenging and may result in low fluency. Based on that, we propose a two-stage searching strategy named \textbf{VCA-NN (nearest neighbours)} to obtain proper source speech (represented as $S_{NN}$ in Figure~\ref{fig:VC}).

\subsubsection{Stage 1} 
We train an SR model $\Phi$ with the classification loss based on all given labelled utterances. As a basic SR model with the distinguishing ability, $\Phi$ can extract the speech features of all utterances in $\mathcal{R}$.

\subsubsection{Stage 2}
Then we calculate the cosine similarity among all utterances from the target set $\mathcal{T} = \{(x_i,y_i)\}_{i=1}^{N}$ and source set $\mathcal{S} = \{(x_j)\}_{j={N+1}}^{|\mathcal{R}|}$ using the extracted features. These features come from the speaker recognition model developed in Stage 1. For each target utterance $(x_i,y_i) \in \mathcal{T}$, we intelligently sample a set $\mathcal{P}_i$ from $\mathcal{S}$. $\mathcal{P}_i$ contains $K$ utterances that exhibit the highest similarity score to the target speech $(x_i,y_i)$, $K$ is the generation coefficient.

\begin{equation}
    \label{e1}
    \mathcal{P}_i = \mathop{\text{arg top-}K} cos(\frac{\Phi(x_i) \cdot \Phi(x_j)}{||\Phi(x_i)|| \cdot ||\Phi(x_j)||})
\end{equation}

where $j \in [N+1, |\mathcal{R}|]$. In other words, for each target speech $(x_i,y_i)$, VCA-NN can obtain $K$ pseudo utterances based on the corresponding set $\mathcal{P}_i$. This approach can achieve higher voiceprint similarity between the source and target speech. By selecting a source speech proximate to the target speech, the generated pseudo speech can be smoother and more fluent, which further boosts SR training.

\section{Dataset creation and experimental settings}
\subsection{Dataset generation}
VoxCeleb2~\cite{Voxceleb2} is the large-scale real-world SR dataset. It contains 1,092,009 speech utterances from 5,994 speakers. Based on VoxCeleb2, we create three training sets to represent the aforementioned scenarios: 1) \textbf{Vox-Semi dataset}: For semi-supervised scenario, Vox-Semi contains the labelled part~(target speech set) and the unlabelled part~(source speech set). The labelled part has 1,000 speakers, each with 10 utterances as target speech. The unlabelled part contains 40,000 utterances as source speech. VCA generates $K=9$ samples for each labelled target speech. The augmented Vox-Semi contains 1,000 speakers, each with 100 utterances. 2) \textbf{Vox-Small dataset}: For small-scale scenario, Vox-Small has 1,000 speakers, while each with 5 utterances. VCA generates $K=9$ samples for each speaker's utterance. This leads to the augmented Vox-Small with 1,000 speakers, each with 50 utterances. 3) \textbf{Vox-Imb dataset}: For imbalanced scenario, Vox-Imb has minority part~(target speech set) and majority part~(source speech set). Majority part has 1000 speakers, each with 10 utterances as source speech, and minority part has 4,000 speakers, each with only 1 utterance as target speech. VCA generates $K=9$ samples for each target speech of minority part. The augmented Vox-Imb contains 5,000 speakers, each with 10 utterances.

\subsection{Implementation details}
During evaluation, the validation set is Vox1-O and the test sets are Vox1-E and Vox1-H. They are the subsets of VoxCeleb1 dataset, which contains speech utterances from 1,251 speakers~\cite{Voxceleb}. All evaluation sets contain several trials, each has two utterances. The cosine similarity between the output features of the given trial is calculated. The metric is the equal error rate (EER) and minDCF (minC), lower is better. The additive data augmentation methods are used in all experiments to boost training~\cite{MUSAN, RIRS}. Diff-VC is trained on the LibriTTS dataset ~\cite{zen2019libritts}. The baseline in our work is to train the speaker model on the original dataset, we also re-implement some previous methods on our datasets with the same experimental settings for comparison.

\section{Results and analysis}
\subsection{Results comparison}
The baseline in our work is to train the speaker model directly on the original datasets without VCA. In Table~\ref{tab:result}, we report the results of the speaker model trained on three datasets that are augmented with our proposed VCA-RS and VCA-NN strategies. In each scenario, we also re-implement several previous methods for comparison.

\subsubsection{Results for semi-supervised learning}
The upper section of Table \ref{tab:result} illustrates the results for the semi-supervised learning scenario (Vox-Semi), where VCA methods outperform the baseline and the semi-supervised contrastive learning solution~\cite{inoue2020semi}. In particular, VCA-NN achieves an average improvement of 5.12\% in EER compared to the baseline for three evaluation sets.

\subsubsection{Results for small-scale learning}
The middle part of Table~\ref{tab:result} reports the performance of the Vox-Small dataset. Here, VCA-RS and VCA-NN perform competitively with the previous speaker augmentation method~\cite{yamamoto19_interspeech}. Due to the limited dataset scale, VCA-NN is not very stable. VCA-RS exhibits the best results with an average deduction of 17.62\% in EER relative to the baseline.

\subsubsection{Results for imbalanced learning}
The bottom portion of Table~\ref{tab:result} compares the performance of VCA against weighted-learning method~\cite{barua2012mwmote} and oversampling-based method~\cite{kotsiantis2006handling} for imbalanced learning. Among these, VCA-NN methods achieve 6.99\% EER and 0.435 minDCF on the Vox1-O dataset, while baseline only achieves 7.80\% EER and 0.467 minDCF.

\subsubsection{Summary} As a unified solution, VCA can effectively enhance the performance of SR model under different scenarios, and the improvements are comparable to those of the previous specific solution. Meanwhile, while there are instances where VCA-RS has better performance, VCA-NN generally outperforms VCA-RS.

\begin{table}[ht] 
\centering
    \vspace{-2mm}
  \caption{The comparison for different augmentation methods under scenarios of semi-supervised learning, small-scale learning, and imbalanced learning.}
  \begin{tabular}{p{1.4cm}<{\centering}p{.7cm}<{\centering}p{.8cm}<{\centering}p{.7cm}<{\centering}p{.8cm}<{\centering}p{.7cm}<{\centering}p{.8cm}<{\centering}} 
    \hline
    \multicolumn{7}{c}{Vox-Semi} \\
    \hline
    \multirow{2}{*}{Method} & \multicolumn{2}{c}{Vox1-O} & \multicolumn{2}{c}{Vox1-E} & \multicolumn{2}{c}{Vox1-H} \\
    &  EER & minC & EER & minC & EER & minC \\
    \hline
    Baseline & $8.90$ & $0.529$& $9.06$& $0.538$& $13.81$& $0.669$ \\ 
    \cite{inoue2020semi}$^\dagger$ & $8.50$ & $0.499$& $8.83$& $0.525$& $13.44$& $0.655$ \\ 
    VCA-RS & $8.21$ & $0.500$& $8.70$& $0.548$& $13.36$& $0.687$ \\ 
    VCA-NN & $8.11$ & $0.476$& $8.74$& $0.533$& $13.38$& $0.675$ \\ 
    \hline
    \hline
    \multicolumn{7}{c}{Vox-Small} \\
    \hline
    \multirow{2}{*}{Method} & \multicolumn{2}{c}{Vox1-O} & \multicolumn{2}{c}{Vox1-E} & \multicolumn{2}{c}{Vox1-H} \\
    &  EER & minC & EER & minC & EER & minC \\
    \hline
    Baseline & $11.89$ & $0.624$ & $12.14$ & $0.673$ & $17.83$ & $0.787$ \\ 
    \cite{yamamoto19_interspeech}$^\dagger$ & $9.85$ & $0.526$ & $10.44$ & $0.596$ & $15.83$ & $0.716$ \\ 
    VCA-RS & $9.51$ & $0.554$ & $9.91$ & $0.591$ & $15.25$ & $0.726$ \\ 
    VCA-NN & $9.95$ & $0.562$ & $10.48$ & $0.612$ & $15.43$ & $0.732$ \\ 
    \hline
    \hline
    \multicolumn{7}{c}{Vox-Imb} \\
    \hline
    \multirow{2}{*}{Method} & \multicolumn{2}{c}{Vox1-O} & \multicolumn{2}{c}{Vox1-E} & \multicolumn{2}{c}{Vox1-H} \\
    &  EER & minC & EER & minC & EER & minC \\
    \hline
    Baseline & $7.80$ & $0.467$& $8.27$& $0.491$& $12.55$& $0.619$ \\ 
    \cite{barua2012mwmote}$^\dagger$ & $7.67$ & $0.453$& $8.00$& $0.475$& $12.18$& $0.597$ \\
    \cite{kotsiantis2006handling}$^\dagger$ & $7.55$ & $0.478$& $8.14$& $0.480$& $12.31$& $0.607$ \\
    VCA-RS & $7.13$ & $0.427$& $7.41$& $0.446$& $11.20$& $0.571$ \\ 
    VCA-NN & $6.99$ & $0.435$& $7.32$& $0.444$& $11.18$& $0.567$ \\ 
    \hline    
    \label{tab:result}
  \end{tabular}
  \vspace{-2mm}
  \begin{tablenotes} \footnotesize
\item $^\dagger$: our re-implementation on these datasets.
\end{tablenotes}
\end{table}

\subsection{Ablation study}

\subsubsection{Robutness for different backbones}
\begin{figure}[!htb]
    \centering
    \includegraphics[width=\linewidth]{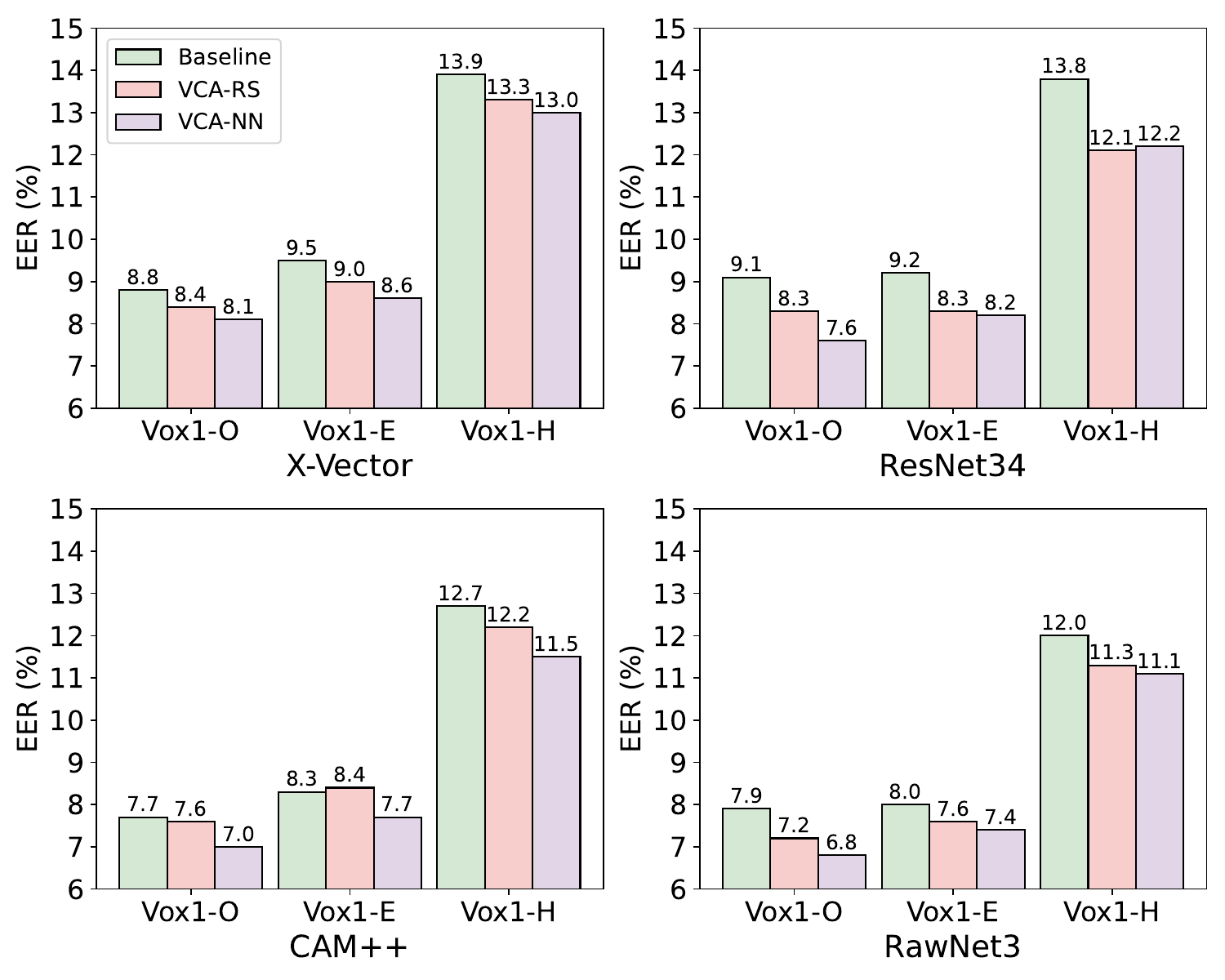}
    \caption{Performance of baseline, VCA-RS and VCA-NN for different backbone architecture trained on the Vox-Imb dataset.}
    \label{fig:backbone}
\end{figure}

Figure~\ref{fig:backbone} investigates the robustness of VCA by analyzing the SR performance of various backbones, including X-Vector~\cite{xvectors}, ResNet34~\cite{he2016deep, chung2020defence}, CAM++~\cite{wang2023cam++}, and RawNet3~\cite{jung22_interspeech} trained on the Vox-Imb. VCA-NN reaches the lowest EER across multiple test sets, with VCA-RS following closely. These results indicate that the integration of VCA methods significantly boosts the performance of SR systems for various backbone architectures.

\subsubsection{Efficiency of VCA with different number of pseudo utterances}
Table~\ref{tab:abl} compares the performance of the VCA-RS and VCA-NN for the different generation coefficients $K$ for each target speech, where $K=0$ denotes the baseline. VCA-NN generally performs better than VCA-RS since it can generate pseudo speech with a high-quality representation. Meanwhile, a small $K$ will even damage the system in some evaluation sets.

\begin{table}[!htb]
\setlength{\tabcolsep}{5.6pt}
\caption{EER (\%) of VCA-RS and VCA-NN for different numbers of generated utterances trained on the Vox-Semi dataset. $K = 0$ denotes the baseline method.}
\begin{tabular}{p{1.8cm}<{\centering}p{1.5cm}<{\centering}p{.8cm}<{\centering}p{.8cm}<{\centering}p{.8cm}<{\centering}p{.8cm}<{\centering}}
\hline
    \multirow{2}{*}{Evaluation set} & \multirow{2}{*}{{Method}} & \multicolumn{4}{c}{{Generation coefficient $K$}} \\
        &   & 0 & 3 & 6 & 9 \\
\hline
\multirow{2}{*}{{Vox1-O}} & VCA-RS & \multirow{2}{*}{8.90}  & 8.74 & 8.64 & 8.21 \\
                          & VCA-NN &                        & 8.55 & 8.24 & 8.11 \\
                                   \hline
\multirow{2}{*}{{Vox1-E}} & VCA-RS & \multirow{2}{*}{9.06}  & 9.30 & 8.89 & 8.70 \\
                          & VCA-NN &                        & 9.11 & 8.62 & 8.74 \\
                                   \hline
\multirow{2}{*}{{Vox1-H}} & VCA-RS & \multirow{2}{*}{13.81}  & 14.16 & 13.61 & 13.36 \\
                          & VCA-NN &                         & 14.04 & 13.51 & 13.38 \\
                                   \hline   
\end{tabular}
\label{tab:abl}
\end{table}
\vspace{3mm}

\section{Conclusion}
In this paper, we propose VCA, a new data augmentation method for defective speaker recognition datasets based on voice conversion. Our experiments prove that VCA can efficiently handle semi-supervised, small-scale, and imbalanced learning scenarios with a unified format. Meanwhile, searching for the source speech with a similar representation to the target speech is an effective strategy. Our VCA provides a new direction to solve the speaker recognition problems from the data aspect. Two points can be further studied for VCA: First, the low-quality pseudo utterances can damage the SR system. Second, due to the speed issue, we have yet to study the feasibility of extending VCA to the large-scale dataset as a general augmentation solution. The proposed technique can also be extended to other areas such as speaker anonymization.

\section{Acknowledgment}
This research is supported by: Shenzhen Science and Technology Program (Shenzhen Key Laboratory, Grant No. ZDSYS20230626091302006); Shenzhen Science and Technology Research Fund (Fundamental Research Key Project, Grant No. JCYJ20220818103001002); Program for Guangdong Introducing Innovative and Entrepreneurial Teams, Grant No. 2023ZT10X044.

\newpage
\printbibliography

\end{document}